\documentclass[aps,preprint]{revtex4}
\usepackage{amssymb}
\usepackage{graphicx}
\usepackage{amsmath}

\begin{document}

\title{Green's function of a finite chain and the discrete Fourier
transform}

\author{S. Cojocaru}
\affiliation{Dipartimento di Fisica ``E.R. Caianiello",
Universit{\`a} degli Studi di Salerno, Italy \\
Institute of Applied Physics, Chi\c{s}in\u{a}u, Moldova}

\begin{abstract}
A new expression for the Green's function of a finite one-dimensional
lattice with nearest neighbor interaction is derived via discrete Fourier
transform. Solution of the Heisenberg spin chain with periodic and open
boundary conditions is considered as an example. Comparison to Bethe ansatz
clarifies the relation between the two approaches.
\end{abstract}

\maketitle

\section{Introduction}

Lattice Green functions have a broad spectrum of applications in physics and
allow to study dynamic and statistical properties. \cite{1,2,3,4} Usually
the lattice is assumed to be infinite while distances remain finite and
discrete. \cite{5,6,7} Often the range of relevant excitations allows to
consider the system as continuous and apply field theoretic methods. \cite{8}
Accordingly, the integral Fourier transform or infinite Fourier series are
commonly used as powerful tools for such problems. It therefore appears
natural to employ the discrete, or finite, Fourier transform (DFT) when the
finite size is relevant. However, DFT has mainly served for numerical
analysis or in fast Fourier transform calculations \cite{9} and much less as
an analytic method. Most of the known exact results on finite systems have
been obtained in the Bethe ansatz approach \cite{10} (see, e.g., \cite{11,12}
for recent review) that is based on a different strategy. In \cite{3} the
Bethe ansatz solution is compared to the solution by Fourier series in the
thermodynamic limit of the periodic quantum spin chain. However it is noted
that the relation between them remains obscure. For instance, the bound
two-magnon excitation is characterized by the same eigenenergy, but only one
eigenstate is found in the Fourier approach, instead of the two solutions in
Bethe ansatz. This discrepancy has been resolved in \cite{13} by an
approximate treatment of finite size corrections. In the present paper it is
shown that the DFT approach recovers the exact solution. The derivation
presents interest in itself, in view of further extensions. As another
example, some results on open boundary conditions obtained by Bethe ansatz
\cite{14} are recovered.

The finite lattice sums are first expressed in terms of infinite Fourier
series. For the nearest neighbor interaction the series represent an
expansion in modified Bessel functions. Its Laplace transform results in
analytic expressions for the finite lattice Green functions and allows to
determine the corresponding discrete Fourier transform.

The standard definitions of the Fourier series for a function periodically
continued from the interval $x\in \left[ 0;2\pi \right) $ are%
\begin{equation}
f\left( z;x\right) =\sum_{n=-\infty }^{\infty }c_{n}\exp \left( inx\right) ,
\label{e1}
\end{equation}%
\[
c_{n}=\frac{1}{2\pi }\int_{0}^{2\pi }f\left( z;y\right) \exp \left(
-iny\right) dy,
\]%
where the parameters $z$ and $\delta $ (below) take real values. The Fourier
expansion of the lattice sum%
\[
\frac{1}{N}\sum_{m=0}^{N-1}f\left( z;\delta +2\pi \frac{m}{N}\right)
\]%
can then be written as%
\[
\sum_{n=-\infty }^{\infty }c_{n}\exp \left( in\delta \right) \left[ \frac{1}{%
N}\sum_{m=0}^{N-1}\exp \left( i2\pi m\frac{n}{N}\right) \right]
=\sum_{k=-\infty }^{\infty }c_{kN}\exp \left( i\delta kN\right) .
\]%
It has been taken into account that the sum in the square brackets above
gives either $1$ or $0,$ provided $n/N$ is an integer ($k$) or not. We note
that the same property is essential for the definition of the DFT of a
function $a\left( X\right) $ defined on a set of $N$ consecutive integers ($%
0,1,..N-1$ ):%
\begin{equation}
a\left( X\right) =\frac{1}{N}\sum_{m=0}^{N-1}b\left( q=\frac{2\pi }{N}%
m\right) \exp \left( iqX\right) .  \label{DFT}
\end{equation}%
Then, with the expression of the Fourier coefficients $c$ in (\ref{e1}), our
lattice sum becomes%
\begin{equation}
\frac{1}{N}\sum_{m=0}^{N-1}f\left( z;\delta +2\pi \frac{m}{N}\right)
=\sum_{k=-\infty }^{\infty }\frac{1}{2\pi }\int_{0}^{2\pi }f\left(
z;y\right) \exp \left( ikN\left( \delta -y\right) \right) dy.  \label{e2}
\end{equation}%
Now the problem of finite summation has been reduced to integration and
series summation.

To motivate a particular choice of $f\left( z;y\right) $ in (\ref{e2}) we
note that dynamics of different models on a finite d-dimensional lattice can
be described in terms of Green functions of a generic form%
\begin{equation}
G\left( p;X,...,Y\right) =\frac{1}{N^{d}}\sum_{Q_{1}}...\sum_{Q_{d}}\frac{%
\exp \left( iQ_{1}X_{1}+...+iQ_{d}Y_{d}\right) }{p-\varepsilon \left(
Q_{d},...,Q_{d}\right) },  \label{e3}
\end{equation}%
where $p$ is the spectral parameter, $\varepsilon $ is the lattice
dispersion and $\left( X_{1},...,\ X_{d}\right) $ are the discrete
coordinates on a periodic lattice $(X\in \left[ 0,1,...,N-1\right] ).$ The
conjugate "quasimomenta" $\left( Q_{1},...,Q_{d}\right) $ defined as
\begin{equation}
Q_{i}=\frac{2\pi }{N}m_{i}+\frac{\Delta }{N};\ m_{i}=0,...,\ N-1,
\label{e30}
\end{equation}%
are already present in the l.h.s. of (\ref{e2}) and their definition is
explained in the example below. For simple lattices the dispersion is a sum
of cosines%
\begin{equation}
\varepsilon =\alpha _{1}\cos Q_{1}+...+\alpha _{d}\cos Q_{d}.  \label{e31}
\end{equation}%
Therefore the Green function can be related to a Laplace transform of
\begin{equation}
f\left( z;y\right) =\exp \left( z\left( \alpha _{1}\cos y_{1}+...+\alpha
_{d}\cos y_{d}\right) \right) .  \label{e4}
\end{equation}

\section{Lattice Green's function}

Let us consider the consequences of (\ref{e2}) for the choice (\ref{e4}) of $%
f\left( z;y\right) $ corresponding to the chain with nearest neighbor
interaction. Integration gives the modified Bessel function $I_{kN}\left(
z\right) $ and (\ref{e2}) becomes%
\begin{equation}
\frac{1}{N}\sum_{m=0}^{N-1}\exp \left( z\cos Q_{m}\right) =\sum_{k=-\infty
}^{\infty }I_{kN}\left( z\right) \exp \left( ik\Delta \right) =I_{0}\left(
z\right) +2\sum_{k=0}^{\infty }I_{kN}\left( z\right) \cos \left( k\Delta
\right) .  \label{e5}
\end{equation}%
The parameters $z$ and $\Delta $ can now be continued to complex values,
e.g., rewriting Eq. (\ref{e5}) in an equivalent form, we obtain a
generalization of the Jacobi expansion~\cite{15}:%
\begin{equation}
\frac{1}{N}\sum_{m=0}^{N-1}\exp \left( \frac{z}{2}\left( w\exp \left( i\frac{%
2\pi m}{N}\right) +\frac{1}{w}\exp \left( -i\frac{2\pi m}{N}\right) \right)
\right) =\sum_{k=-\infty }^{\infty }I_{kN}\left( z\right) w^{kN}.  \label{e6}
\end{equation}%
One can check that the expressions 9.6.33-9.6.40 in \cite{16} are particular
cases of (\ref{e5}) or (\ref{e6}) for $N=1,2.$ For instance, taking $\Delta
=\theta +\frac{\pi }{2}$ in (\ref{e5}) one finds

\begin{equation}
\frac{1}{N}\sum_{m=0}^{N-1}\exp \left( z\cos \left( \frac{\theta }{N}+\frac{%
\pi }{2N}+2\pi \frac{m}{N}\right) \right) =\sum_{k=-\infty }^{\infty
}I_{kN}\left( z\right) \exp \left( i\frac{\pi k}{2}\right) \exp \left(
i\theta k\right)  \label{e61}
\end{equation}%
\[
=I_{0}\left( z\right) +2\sum_{k=1}^{\infty }\left( -1\right)
^{k}I_{2kN}\left( z\right) \cos \left( 2k\theta \right) +2\sum_{k=0}^{\infty
}\left( -1\right) ^{k}I_{\left( 2k+1\right) N}\left( z\right) \sin \left(
\left( 2k+1\right) \theta \right) .
\]%
It is also not difficult to obtain a further generalization of these results
by including the space dependent factor of the Green function (\ref{e3}) and
a dispersion of the form:
\begin{equation}
\varepsilon =b\cos Q+c\sin Q,  \label{e62}
\end{equation}%
so that the propagator can be written as
\begin{equation}
G\left( p;X\right) =\frac{1}{N}\sum_{m=0}^{N-1}\frac{\exp \left( iQX\right)
}{p-\cos \left( Q+\eta \right) }.  \label{e7}
\end{equation}%
It corresponds to the Laplace transform of the lattice sum
\begin{equation}
L=\frac{1}{N}\sum_{m}\exp \left( z\cos \left( \frac{\Delta }{N}+2\pi \frac{m%
}{N}+\eta \right) \right) \exp \left( i\left( \frac{\Delta }{N}+2\pi \frac{m%
}{N}\right) X\right).  \label{e8}
\end{equation}%
By denoting $\xi \equiv \Delta +\eta N$ we have
\begin{equation}
L=\exp \left( -i\eta X\right) \frac{1}{N}\sum_{m=0}^{N-1}\exp \left( z\cos
\left( \frac{\xi }{N}+2\pi \frac{m}{N}\right) \right) \exp \left( i\left(
\frac{\xi }{N}+2\pi \frac{m}{N}\right) X\right) .  \label{e81}
\end{equation}%
Then substitute (\ref{e81}) into (\ref{e2})
\[
L=\exp \left( -i\eta X\right) \sum_{k=-\infty }^{\infty }\exp \left( ik\xi
\right) \frac{1}{2\pi }\int_{0}^{2\pi }\exp \left( z\cos y\right) \exp
\left( -i\left( kN-X\right) y\right) dy,
\]%
and finally obtain%
\begin{eqnarray}
&&\frac{1}{N}\sum_{m=0}^{N-1}\exp \left( z\cos \left( \frac{\Delta }{N}+2\pi
\frac{m}{N}+\eta \right) \right) \exp \left( i\left( \frac{\Delta }{N}+2\pi
\frac{m}{N}\right) X\right) \\
&=&\exp \left( -i\eta X\right) \sum_{k=-\infty }^{\infty }I_{kN+X}\left(
z\right) \exp \left( -ik\left( \Delta +\eta N\right) \right) .  \label{e91}
\end{eqnarray}%
The above expressions demonstrate a direct relation of the generalized
Jacobi expansion to the finite lattice problem. However their main advantage
becomes clear by realizing that the increasing orders of the Bessel function
in these series correspond to the consecutive terms of the asymptotic $N-$%
expansion of the lattice propagators. I.e., the term with $k=0$ describes
the thermodynamic limit ( $N\longrightarrow \infty $ ) and the first finite-$%
N$ correction is contained in the term with $k=1.$ One can also evaluate the
convergence of these series, that depends on lattice and on the spectral
parameter $p$. We will provide specific examples and explain the choice of
"quasimomentum" $Q$ and its apparent conflict with periodicity of the
lattice.

The lattice Green function (\ref{e7}) can now be obtained from the Laplace
transform of (\ref{e91}) with respect to $z$ \cite{17}%
\[
\int_{0}^{\infty }\exp \left( -pz\right) I_{n}\left( z\right) dz=\frac{\exp
\left( -\left\vert n\right\vert v\right) }{\sinh v},
\]%
where $n$ is an integer and we have introduced the notation $p\equiv \cosh v$
( convergence of the Laplace transform requires Re $p>1$ ). Thus
\[
\frac{\sinh v}{N}\sum_{m=0}^{N-1}\frac{\exp \left( i\left( \frac{\Delta }{N}%
+2\pi \frac{m}{N}\right) X\right) }{\cosh v-\cos \left( \frac{\Delta }{N}%
+2\pi \frac{m}{N}+\eta \right) }
\]%
\[
=\sum_{k=-\infty }^{\infty }\exp \left( -v\left\vert kN+X\right\vert \right)
\exp \left( -ik\left( \Delta +\eta N\right) -i\eta X\right) ,
\]%
leads to the following expression for the propagator (\ref{e7}):%
\[
\frac{1}{N}\sum_{m=0}^{N-1}\frac{\exp \left( i\left( \frac{\Delta }{N}+2\pi
\frac{m}{N}\right) X\right) }{\cosh v-\cos \left( \frac{\Delta }{N}+2\pi
\frac{m}{N}+\eta \right) }=\frac{\exp \left( i\eta \left( N/2-X_{M}\right)
+i\left( M+1/2\right) \Delta \right) }{2\sinh v}
\]%
\begin{equation}
\times \left( \frac{\exp \left( v\left( N/2-X_{M}\right) \right) }{\sinh
\left( \left( vN+i\left( \Delta +\eta N\right) \right) /2\right) }+\frac{%
\exp \left( -v\left( N/2-X_{M}\right) \right) }{\sinh \left( \left(
vN-i\left( \Delta +\eta N\right) \right) /2\right) }\right) .  \label{G}
\end{equation}%
The integer $M$ in the (\ref{G}) defines a translation relating an arbitrary
$X$ to $X_{M}$ from the \textit{main} interval$:$%
\begin{equation}
X_{M}=0,1,...N-1,  \label{g0}
\end{equation}%
\[
X=X_{M}+MN.
\]%
We note the transformation properties following from the above definition,%
\[
X\rightarrow N+X:\ \ \ \ X_{M}\rightarrow X_{M},\ \ \ \ M\rightarrow M+1;
\]%
\[
X\rightarrow -X:\ \ \ \ \ X_{M}\rightarrow N-X_{M},\ \ \ \ M\rightarrow
-\left( M+1\right) ;
\]%
\begin{equation}
X\rightarrow N-X:\ \ \ \ \ X_{M}\rightarrow N-X_{M},\ \ \ \ M\rightarrow -M.
\label{transform}
\end{equation}%
After the r.h.s. of (\ref{G}) has been found, it is now easy to verify that
the inverse DFT (i.e. taking the summation in $X$ ) indeed reproduces the
function in the l.h.s.With (\ref{transform}) one also finds another useful
form of DFT (\ref{G}):%
\[
\frac{2\sinh v}{N}\sum_{m=0}^{N-1}\frac{\cos \left( \left( \frac{\Delta }{N}%
+2\pi \frac{m}{N}\right) X\right) }{\cosh v-\cos \left( \frac{\Delta }{N}%
+2\pi \frac{m}{N}+\eta \right) }=\frac{\cosh \left( \left( v+i\eta \right)
\left( \frac{N}{2}-X_{M}\right) +i\left( M+\frac{1}{2}\right) \Delta \right)
}{\sinh \left( \frac{N}{2}\left( v+i\eta \right) +i\frac{\Delta }{2}\right) }
\]%
\[
+\frac{\cosh \left( \left( v-i\eta \right) \left( \frac{N}{2}-X_{M}\right)
-i\left( M+\frac{1}{2}\right) \Delta \right) }{\sinh \left( \frac{N}{2}%
\left( v-i\eta \right) -i\frac{\Delta }{2}\right) }.
\]%
\[
\frac{2i\sinh v}{N}\sum_{m=0}^{N-1}\frac{\sin \left( \left( \frac{\Delta }{N}%
+2\pi \frac{m}{N}\right) X\right) }{\cosh v-\cos \left( \frac{\Delta }{N}%
+\eta +2\pi \frac{m}{N}\right) }=\frac{\sinh \left( \left( v+i\eta \right)
\left( \frac{N}{2}-X_{M}\right) +i\left( M+\frac{1}{2}\right) \Delta \right)
}{\sinh \left( \left( v+i\eta \right) \frac{N}{2}+i\Delta /2\right) }
\]%
\begin{equation}
-\frac{\sinh \left( \left( v-i\eta \right) \left( \frac{N}{2}-X_{M}\right)
-i\left( M+\frac{1}{2}\right) \Delta \right) }{\sinh \left( \left( v-i\eta
\right) \frac{N}{2}-i\Delta /2\right) }.  \label{G1}
\end{equation}

The condition $\cosh v\geqslant 1$ in our derivation was due to the use of
the Laplace transform. It is however important to mention that the above
expressions can be analytically continued to arbitrary complex values of the
parameters ( $v,\eta ,\Delta $ ), since the resonance conditions (zeros of
the denominator in the l.h.s.) correspond to simple poles. For instance, a
physically important possibility is the region Re $p<1$ ($p\neq \cos Q$),
where respective solutions of the Schr\"{o}dinger equation with Im $p=0$ are
known as scattered states.

\section{The Heisenberg ring}

The model of a cyclic chain of $N$ quantum spins $S=1/2$ with ferromagnetic
nearest neighbor interaction was the first to be solved by Bethe ansatz \cite%
{10} and it offers a clear test for the above formal results. The two-magnon
subspace of the eigenfunctions of the Hamiltonian
\begin{equation}
H=-J\sum_{\left\langle ij\right\rangle }^{N}\mathbf{S}_{i}\cdot \mathbf{S}%
_{j},  \label{f1}
\end{equation}%
contains the main features of the interaction problem and is described by
the amplitude of two spins flipped on the sites $n_{1}$and $n_{2}:$
\[
|\psi \rangle =\!\sum_{1\leq n_{1}<n_{2}\leq
N}A(n_{1},n_{2})S_{n_{1}}^{-}S_{n_{2}}^{-}|0\rangle .
\]%
Within the Bethe ansatz approach a phase shift parameter is introduced to
account for nearest neighbor interaction in terms of a boundary condition
problem. I.e., by matching this free parameter in the noninteracting
particle form of the ansatz wave function at large separation to satisfy the
Schrodinger equation when the two overturned spins are nearest neighbors.
\cite{11} An alternative approach is based on a representation of the
Schrodinger equation in a form consisting of two contributions: free motion
and interaction terms. \cite{3}
\begin{equation}
\left[ E-2J\right] a(X)+J\;\cos \left( \frac{P}{2}\right) \left(
a(X+1)+a(X-1)\right) =  \label{1}
\end{equation}%
\[
J\left[ \cos \left( \frac{PX}{2}\right) a(0)-a(X)\right] \left( \delta
_{X,\;1}+\delta _{X,\;\;N-1}\right) ,
\]%
where $a(X)$ represents the amplitude of the relative position $X$ of the
flipped spins, $E$ is the excitation energy over the ground state with all
spins parallel and $P$ is the total momentum associated to translation
symmetry.%
\begin{eqnarray}
A(n_{1},n_{2}) &=&\frac{\exp (iPR)}{\sqrt{N}}\;a(X)\;;  \label{general} \\
R &=&\frac{n_{1}+n_{2}}{2};\ \ \ X=(n_{2}-n_{1})=1,...,N-1\;.  \label{g1}
\end{eqnarray}%
The split form of (\ref{1}) is convenient to set up a perturbation theory
and is widely used in condensed matter, when the exact solution is not
known. Then the solution is sought in terms of Fourier transform. The
negative values of $X$ correspond to transposition of the two overturned
spins, i.e., to the same state. At the same time it is a common practice to
chose the interval $X$ symmetrically with respect to zero ($-N/2,\ \left(
N-1\right) /2$ ) and interpreting negative values as "left neighbor" and
positive values as "right neighbor". The respective amplitudes do not
necessarily coincide. Therefore here the definition in (\ref{g0}) is
preferred to avoid this ambiguity and for a straightforward comparison to
Bethe ansatz that follows the same counting convention. For instance, if the
"right neighbor" is defined by $X$ , then the "left" one corresponds to $N-X.
$

First we note that quantization of the total momentum $P=2\pi k/N,$ $%
k=0,1,..,N-1$ in (\ref{general}) follows from translation of the two spin
complex as a whole $A(n_{1},n_{2})=A(n_{1}+N,n_{2}+N).$ Respectively, the
quasimomenta $Q$ in (\ref{e7}), that are conjugate to relative distance $X,$
should be determined by the boundary conditions on the displacement of
separate spin flips. Thus the amplitude of relative motion in (\ref{general}%
) should satisfy the condition%
\begin{equation}
a\left( X\right) =\exp \left( i\pi k\right) a\left( X+N\right) ,  \label{33}
\end{equation}%
following from the translation $A(n_{1},n_{2})=A(n_{1},n_{2}+N).$ According
to (\ref{33}) the relative amplitude is periodic either on a length of the
chain or on a double length, $2N,$ depending on the parity of the total
momentum quantum number $k=PN/2\pi .$ In the latter case the amplitude
changes sign after completing the first cycle and may be denoted as
antisymmetric, $a,$ to distinguish from the former, symmetric mode $s$,
which completes the period after the first cycle. Together with the property
$a\left( X\right) =a\left( -X\right) $ due to spin transposition symmetry,
the above relations define the continuation of the solutions of (\ref{1})
from the main interval (\ref{g1}) to arbitrary values of $X$. By applying
these relations to the Fourier expansion%
\begin{equation}
a(X)=\frac{1}{N}\sum_{Q}b(Q)\cos \left( QX\right) ,  \label{F}
\end{equation}%
we obtain the condition on $Q$%
\[
\exp \left( iN\left( \frac{P}{2}\pm Q\right) \right) ~=1.
\]%
It determines the two sequences corresponding to the two types of modes
\[
Q_{s}=\frac{2\pi l}{N}\;;\;l=0,1,...,N-1,
\]%
\begin{equation}
Q_{a}=\frac{2\pi l}{N}+\frac{\pi }{N}\;;\;l=0,1,...,N-1.  \label{g2}
\end{equation}%
These quasimomenta correspond to $\eta =0$ and $\Delta _{s}=0,\ \Delta
_{a}=\pi $ in the general expressions (\ref{e30}). The state counting
convention for a one-dimensional lattice is to require that the second
argument in the amplitude $A(n_{1},n_{2})$ is larger then the first. This
leads to an additional relation $A(n_{1},n_{2})=A(n_{2},n_{1}+N)$ imposing a
constraint on the amplitude for the main interval%
\begin{equation}
a\left( X\right) =\exp \left( i\pi k\right) a\left( N-X\right) .  \label{34}
\end{equation}%
For instance,%
\begin{equation}
a_{a}\left( X\right) =-a_{a}\left( N-X\right) =-a_{a}\left( N+X\right)
=a_{a}\left( -X\right) .  \label{35}
\end{equation}%
One can check that (\ref{34}) is automatically satisfied for the amplitude (%
\ref{F}) and (\ref{g2}).

Solving (\ref{1}) for the $b\left( Q\right) $ in (\ref{F}), we find%
\[
b(Q)=C\left( P\right) \frac{\cos \left( Q\right) }{\cosh v-\cos Q},
\]%
where the constant $C\left( P\right) $ is
\[
C=-\frac{1}{N}\sum_{Q^{^{\prime }}}b(Q^{^{\prime }})\left[ \cos \left( \frac{%
P}{2}\right) -\cos Q^{^{\prime }}\right] .
\]%
The parameter $v$ is related to the eigenenergy $E=2J\left( 1-\cos \left(
P/2\right) \cosh v\right) $ and is determined by the compatibility equation
\[
1=\frac{1}{N\cos \left( \frac{P}{2}\right) }\sum_{Q^{^{\prime }}}\frac{\cos
\left( Q^{^{\prime }}\right) \left[ \cos Q^{^{\prime }}-\cos \left( \frac{P}{%
2}\right) \right] }{\cosh v-\cos Q^{^{\prime }}}.
\]%
From the DFT formulas (\ref{G}), (\ref{G1}) and (\ref{g2}) we find the
eigenenergy equation for each type of eigenstate. For instance,%
\[
\frac{1}{N}\sum_{Q_{a}}\frac{\cos \left( Q_{a}\right) }{\cosh v_{a}-\cos
Q_{a}}=\frac{1}{2}\frac{\sinh \left( v_{a}\left( N-2\right) /2\right) }{%
\sinh \left( v_{a}\right) \cosh \left( v_{a}N/2\right) }.
\]%
After simple algebra we finally obtain%
\begin{equation}
\left( \cosh v_{a,s}-\cos \left( \frac{P}{2}\right) \right) \coth \left(
\frac{Nv_{a,s}-i\Delta _{a,s}}{2}\right) =\sinh v_{a,s}.  \label{36}
\end{equation}%
Eqs. (\ref{36}) indeed coincide with the Bethe ansatz result \cite{10,11}.
Solutions with real values of $v$ ,\ describing the two bound states
mentioned earlier, merge in the thermodynamic limit to: $v=-\ln \left( \cos
P/2\right) $, $E=J\sin ^{2}\left( P/2\right) .$ At any given $P$ equation (%
\ref{36}) has a number $\sim N$ of solutions with imaginary values of $v,$
or Re $p<1,$ which correspond to scattered states. These solutions are
characterized by an oscillating space dependence, as is clearly seen from (%
\ref{G1}) (see also (\ref{amp}) below), unlike the smoothly decaying
dependence of the bound states. At finite $N$ they form two distinct classes
of energy bands, shifted with respect to each other by $\sim J/N.$ All the
solutions can be cast in a unified form by using the relation of $\Delta $
to the total momentum ($\Delta =PN/2$ ):
\begin{equation}
a\left( X\right) \sim \cosh \left( v\left( \frac{N}{2}-\left\vert
X\right\vert \right) +i\frac{PN}{4}\right) ,  \label{amp}
\end{equation}%
where $v=v\left( P\right) $ corresponds to a particular solution of (\ref{36}%
) at fixed total momentum $P.$ One can see that the phase shift of $\pi /2$
( stemming from $PN/4$ ) between nearest values of $P$ survives in the
thermodynamic limit. This distinction in symmetry of eigenstates was missed
in the previous Fourier series analysis \cite{3,18} because the effect is
contained in the phase shifts of quasimomenta $Q$ (\ref{g2}) that vanishes
in the thermodynamic limit.

\section{Open boundary conditions}

To consider another illustration of the formalism derived above we will
include, according to \cite{14}, also arbitrary end fields ($\mu $ and $\nu $%
) into the anisotropic Hamiltonian on a chain of length $L$ ( $i,j=0,...,L-1$
). Only a single magnon excitation will be considered, as the interaction is
already present due to noncyclic boundary conditions:%
\[
H=-\sum_{i=0}^{L-2}\left( \eta S_{i}^{z}S_{i+1}^{z}+\frac{1}{2}\left(
S_{i}^{+}S_{i+1}^{-}+S_{i}^{-}S_{i+1}^{+}\right) +\mu S_{0}^{z}+\nu
S_{L-1}^{z}\right) .
\]%
The parameters are dimensionless, i.e., scaled with the exchange constant $%
J_{x}$, and the energy of the state with all spins parallel is $E_{0}=-\eta
\left( L-1\right) /2-\left( \mu +\nu \right) /2.$ The Schrodinger equation
is represented in the form similar to (\ref{1}) with the "interaction" on
the r.h.s.:%
\[
\left( E-E_{0}\right) a\left( j\right) -\left[ \eta a\left( j\right) -\frac{1%
}{2}a\left( j-1\right) -\frac{1}{2}a\left( j+1\right) \right]
\]%
\[
=\delta _{j=L-1}\left( \frac{1}{2}a\left( L\right) -a\left( L-1\right)
\left( \frac{1}{2}\eta -\nu \right) \right)
\]%
\begin{equation}
+\delta _{j=0}\left( \frac{1}{2}a\left( -1\right) -a\left( 0\right) \left(
\frac{1}{2}\eta -\mu \right) \right) .  \label{b1}
\end{equation}%
As can be seen, the nonphysical amplitudes $a\left( -1\right) $ and $a\left(
L\right) $ cancel each other and in the DFT approach are in fact defined by
the Eq. (\ref{b1}) and the transformation properties (\ref{transform}). For
instance, these amplitudes may differ from Bethe ansatz solution, as is the
case of the Heisenberg ring. The two approaches, however, give the same
solution for the physical interval. In DFT we need to consider an interval
in $X$ which can be periodically continued. However unlike the cyclic chain,
here the last spin in the open chain is not at the same time the "left"
nearest neighbor of the first spin, i.e., there can be no periodicity
transformation: $X\rightarrow X+L.$ Instead, our chain can be viewed as a
half of the cyclic chain (ring) of length $N=2L$ ( $j=0,..,2L-1$) where the
two links ($2L-1,0$) and ($L-1,L$) are broken. With this definition we can
apply the DFT formulas (\ref{DFT}) and (\ref{G}) to solve (\ref{b1}). IN
this case $\Delta =\eta =0$ and the quasimomentum is obviously defined as $%
Q=2\pi m/N,$ $m=0,...,2L-1.$ An immediate consequence of transformation
properties (\ref{transform}) is that $a\left( -1\right) =a\left( 1\right) ,$
while $a\left( L\right) $ has to be determined from (\ref{b1}). The Fourier
amplitude resulting from Schrodinger equation is%
\begin{equation}
b\left( Q\right) =\frac{-C_{1}}{\left( \cosh v-\cos Q\right) }-C_{2}\frac{%
\exp \left( -iQ\left( L-1\right) \right) }{\left( \cosh v-\cos Q\right) },
\label{b2}
\end{equation}%
where $\cosh v\equiv \eta -\left( E-E_{0}\right) $ and the two constants are
\[
C_{1}=\frac{1}{2}a\left( -1\right) -a\left( 0\right) \left( \frac{1}{2}\eta
-\mu \right) ,
\]%
\[
C_{2}=\frac{1}{2}a\left( L\right) -a\left( L-1\right) \left( \frac{1}{2}\eta
-\nu \right) .
\]%
Substituting the amplitude $b\left( Q\right) $ into the definition of the
constants one obtains a linear homogeneous system of equations%
\[
C_{1}\left( 1+\frac{1}{2N}\sum_{m=0}^{N-1}\frac{\left[ \exp \left(
-iq\right) -\left( \eta -2\mu \right) \right] }{\cosh v-\cos q}\right)
\]%
\[
=-C_{2}\frac{1}{2N}\sum_{m=0}^{N-1}\frac{\exp \left( -iq\left( L-1\right)
\right) \left[ \exp \left( -iq\right) -\left( \eta -2\mu \right) \right] }{%
\cosh v-\cos q},
\]

\begin{equation}
C_{2}\left( 1+\frac{1}{2N}\sum_{m=0}^{N-1}\frac{\left[ \exp \left( iq\right)
-\left( \eta -2\nu \right) \right] }{\cosh v-\cos q}\right)   \label{b3}
\end{equation}%
\[
=-C_{1}\frac{1}{2N}\sum_{m=0}^{N-1}\frac{\exp \left( iq\left( L-1\right)
\right) \left[ \exp \left( iq\right) -\left( \eta -2\nu \right) \right] }{%
\cosh v-\cos q}.
\]%
which defines the eigenenergies and eigenfunctions of the equation (\ref{b1}%
). From (\ref{G}) we have%
\[
\frac{1}{N}\sum_{m=0}^{N-1}\frac{\exp \left( -iq\right) }{\cosh v-\cos q}=%
\frac{\cosh \left( v\left( L-1\right) \right) }{\sinh v\sinh \left(
vL\right) },
\]%
\[
\frac{1}{N}\sum_{m=0}^{N-1}\frac{1}{\cosh v-\cos q}=\frac{\cosh \left(
vL\right) }{\sinh v\sinh \left( vL\right) },
\]%
\[
\frac{1}{N}\sum_{m=0}^{N-1}\frac{\exp \left( -iq\left( L-1\right) \right) }{%
\cosh v-\cos q}=\frac{\cosh \left( v\right) }{\sinh v\sinh \left( vL\right) }%
,
\]%
\[
\frac{1}{N}\sum_{m=0}^{N-1}\frac{\exp \left( -iq\left( L-2\right) \right) }{%
\cosh v-\cos q}=\frac{\cosh \left( 2v\right) }{\sinh v\sinh \left( vL\right)
}.
\]%
With these expressions we obtain from (\ref{b3}) the eigenenergy equation of
\cite{14}:%
\begin{equation}
e^{2v\left( L-1\right) }=\frac{\left( e^{-v}-\left( \eta -2\mu \right)
\right) \left( e^{-v}-\left( \eta -2\nu \right) \right) }{\left(
e^{v}-\left( \eta -2\mu \right) \right) \left( e^{v}-\left( \eta -2\nu
\right) \right) }.  \label{b4}
\end{equation}%
The equation is invariant under $v\leftrightarrow -v$ or $\mu
\leftrightarrow \nu .$ By taking the ratio $C_{1}/C_{2}$ from (\ref{b3}) one
obtains from (\ref{G}) the space dependence of the amplitude:%
\begin{equation}
a\left( X\right) =\gamma \left[ \sinh \left( v\left( x+1\right) \right)
-\left( \eta -2\mu \right) \sinh \left( vx\right) \right] .  \label{b5}
\end{equation}
The factor $\gamma =\gamma \left( v,L\right) $ can be found from
normalization. It is now easy to checked that (\ref{b5}) is equivalent to
Bethe ansatz result provided $v$ satisfies the eigenenergy equation (\ref{b4}%
). Interchanging the end fields $\mu \leftrightarrow \nu $ induces an
obvious transformation (up to a phase factor) $a\left( X\right) \rightarrow
a\left( L-1-X\right) $ which amounts to changing the direction of counting
the sites. As noted in, \cite{14} the solution is particularly simple for
the condition
\[
\left( \eta -2\mu \right) \left( \eta -2\nu \right) =1,
\]%
when one gets $e^{-2v}$ in the r.h.s. of (\ref{b4}) and consequently:%
\[
v=i\frac{\pi m}{L},\ m=0,...,2L-1.
\]%
The wave function represents a superposition of two harmonic waves. Looking
at the energy of the excitation
\[
E-E_{0}=\eta -\cos p,
\]%
one can immediately see that in the case of anisotropic ferromagnetic
interaction with $\eta <1$ ($J^{z}<J^{x}$) the ground state is unstable
against long wavelength magnons. The harmonic (purely oscillating) solution
is also obtained in the limit of strong end fields $\mu ,\nu \rightarrow
\infty $ (equivalent to fixed boundaries) since (\ref{b4}) becomes%
\[
e^{2v\left( L-1\right) }=1:v=ip,\ p=\frac{\pi m}{L-1},m=0,...,2L-3.
\]%
Due to the symmetry of the chain mentioned above, only half of solutions is
linearly independent. An additional solution corresponds to bound state
recovering the total number of modes $L$. For an arbitrary choice of
parameters the solution will consist of scattered waves ( Re$v=0$) and bound
states ( Im$v=0$ ).

\section{Conclusions}

A new expression for the finite 1D lattice Green function has been derived
by the discrete Fourier transform approach, by establishing a generalization
of the Jacobi expansion. We note its qualitative advantage compared to the
approach based in infinite Fourier series that it make accessible a new
region of parameters, not available in the thermodynamic limit, but having
an important physical meaning (e.g., scattering wave solutions). In the
Bethe ansatz approach the phase $\theta $ is introduced into the assumed
form of the wave function and is then determined by substitution into the
Schrodinger equation. As a result, the Bethe phase becomes a function of
momentum (and consequently of energy) and contains all the information on
the interaction in the system. In DFT the sequence is just opposite: no
assumption is made on the wave function. The eigenenergy (Schrodinger)
equation is first solved in the momentum space and then the wave function is
obtained via Fourier transform into the direct lattice space. Then the
identification of the Bethe phase is achieved by comparing the two
functions, $\theta =ivN+\pi m.$ The DFT allows to describe the scattering
states on the same footing as the bound states, and at the same time to keep
a physically transparent structure discussed above. It is known that the
bound states are well separated from continuum of scattered states in the
thermodynamic limit of (\ref{f1}), \cite{18} while the description of
scattered states requires special treatment in a Lipmann-Schwinger type
approach \cite{3}. In the DFT approach it becomes clear that the qualitative
change of the wave function is controlled by the effective magnon-magnon
interaction described by the parameter $v$ (e.g., the r.h.s. of Eq. (\ref{1})
): the bound states are formed when this interaction is attractive, while
the scattered states are due to repulsive effective interaction (see \cite%
{13}). One can show that the critical line separating the two classes of
states corresponds to the vanishing of interaction (i.e. the magnons are
free if their dispersion crosses this line) and coincides with the lower
boundary of the continuum of scattered states. The wave function of the
antisymmetric solutions has nodes on the direct lattice and therefore the
corresponding states are more loosely bound. At sufficiently long wavelength
of the excitation the bound states are very close to scattered states and
the antisymmetric ones can indeed become unstable, i.e., decay into
scattered states. This mechanism is responsible for the so-called non-string
behavior of the solutions of the Bethe equations at finite $N.$ \cite{19}
Thus also the physical interpretation of the Bethe solutions becomes more
transparent. The same approach allows to reproduce the BA solution for an
open chain, which is amenable to Fourier expansion by doubling the length of
the chain. This can be viewed as a cyclic ring with broken links (or
impurities) introducing a scattering of magnon excitations and decay of the
wave function with distance. The approach described in the paper presents
interest for further development, since by its construction it is not
limited to one space dimension.

\end{document}